\documentstyle[preprint,epsf,color]{jpsj}

\newcommand{\red}{\color{black}}

\newcommand{\blue}{\color{black}}
\newcommand{\magenta}{\color{black}}

\def \dag{\dagger}

\def \V2c{V_{\rm 2c}}

\def \lb{\overline{l}}
\def \lb0{\overline{l_0}}
\def\ggs{\buildrel\textstyle > \over {\hbox{\raise0.2ex\hbox{$\sim$}}}}
\def\lls{\buildrel\textstyle < \over {\hbox{\raise0.2ex\hbox{$\sim$}}}}
\def\gsim{\,\lower0.75ex\hbox{$\ggs$}\,}
\def\lsim{\,\lower0.75ex\hbox{$\lls$}\,}

\newcommand{\bone}{\sigma_0}

\newcommand{\sigmab}{\mbox{\boldmath $\sigma $}}

\newcommand{\bk}{{\bf k}}
\newcommand{\br}{{\bf r}}

\newcommand{\bw}{{\bf w}}

\newcommand{\bR}{{\bf R}}

\newcommand{\wtilde}{\tilde{w}}

\newcommand{\beq}{\begin{equation}}
\newcommand{\beqn}{\begin{eqnarray}}
\newcommand{\eeq}{\end{equation}}
\newcommand{\eeqn}{\end{eqnarray}}

\newcommand{\YS}[1]{\textcolor{black}{#1}}

\def\runtitle{\YS{Tilted-Cone}-induced easy-plane pseudo-spin ferromagnet and Kosterlitz-Thouless transition in massless Dirac fermions}
\def\runauthor{Akito {\sc  KOBAYASHI}$^{1,2}$, Yoshikazu {\sc  SUZUMURA}$^2$, Hidetoshi {\sc FUKUYAMA}$^3$, 
and Mark O. {\sc GOERBIG}$^4$}

\title{{\magenta \YS{Tilted-Cone}-induced easy-plane pseudo-spin ferromagnet and Kosterlitz-Thouless transition 
in massless Dirac fermions}}

\author{Akito {\sc  KOBAYASHI}$^{1,2}$, Yoshikazu {\sc  SUZUMURA}$^2$, Hidetoshi {\sc FUKUYAMA}$^3$, 
and Mark O. {\sc GOERBIG}$^4$}

\inst{
$^1$Institute for Advanced Research, Nagoya University, Furo-cho, Chikusa-ku, Nagoya, 464-8602 Japan 
 \\ $^2$Department of Physics, Nagoya University, Furo-cho, Chikusa-ku, Nagoya, 464-8602 Japan
 \\ $^3$Department of Applied Physics, Tokyo University of Science, 1-3, Kagurazaka, Shinjuku-ku, Tokyo 162-8601 Japan
 \\ $^4$Laboratoire de Physique des Solides, CNRS UMR 8502, Universite Paris-Sud, F-91405 Orsay Cedex, France
}

\abst{
The possible quantum Hall ferromagnet at a filling factor $\nu =0$ is investigated 
for the zero-energy ($N=0$) Landau level of the two dimensional massless Dirac fermions 
in $\alpha$-(BEDT-TTF)$_2$I$_3$ under pressure with tilted cones and a twofold valley degeneracy 
resulting from time-reversal symmetry.
In the case of the Dirac cones without tilting, the long-range Coulomb interaction in the $N=0$ Landau level
exhibits the $SU(2)$ valley-pseudo-spin symmetry even to the order $O(a/l_{\rm H})$, 
in contrast to $N \ne 0$ Landau levels, where $a$ and $l_{\rm H}$ represent the lattice constant 
and the magnetic length, respectively. 
Such a characteristic comes from a fact that zero-energy states in a particular valley are restricted 
to only one of the spinor components, whereas the other spinor component is necessarily zero. 
In the case of the tilted Dirac cones as found in $\alpha$-(BEDT-TTF)$_2$I$_3$, 
one obtains a non-zero value of the second component and then 
the backscattering processes between valleys becomes non-zero.
It is shown that this fact can lead to easy-plane pseudospin ferromagnetism (XY-type).
In this case, the phase fluctuations of the order parameters can be described 
by the XY model 
leading to Kosterlitz-Thouless transition at lower temperature. 
In view of these theoretical results, experimental findings in resistivity 
of $\alpha$-(BEDT-TTF)$_2$I$_3$ are discussed.}

\recdate{\hspace{35mm}}

\kword
{
massless Dirac fermion, Weyl equation, zero-gap state, $\alpha$-(BEDT-TTF)$_2$I$_3$, graphene, magnetic field, Landau level, electron correlation, quantum Hall ferromagnet, Kosterlitz-Thouless transition
}

\begin{document}

\newcommand{\vct}[1]{\mbox{\boldmath\(#1\)}}
\newcommand{\vep}{\varepsilon}
\newcommand{\dsum}[2]{\displaystyle{\sum_{#1}^{#2}}}
\newcommand{\dint}[2]{\displaystyle{\int_{#1}^{#2}}}
\newcommand{\lrangle}[1]{\langle{#1}\rangle}

\sloppy
\maketitle

\section{Introduction}

The massless Dirac fermions in the quasi-two-dimensional organic conductor 
$\alpha$-(BEDT-TTF)$_2$I$_3$,\cite{Katayama2006ZGS} 
which obey the tilted Weyl equation\cite{Kobayashi2007,Goerbig2008}, have attracted much interest 
because of the mysteries of the experimental findings such as the weak temperature ($T$) dependence of resistivity 
(close to $h/e^2$), the strong $T$-dependence of the Hall coefficient,
\cite{KajitaFirst,Tajima2006JPSJ,Tajima2009PRL,Kobayashi2008} 
and the two-step increase of resistivity with decreasing $T$ in the presence of the magnetic field.
\cite{Tajima2006JPSJ} 

In the absence of magnetic field and under pressure $P=18$kbar, 
the in-plane resistivity decreases weakly with decreasing $T$ from the room temperature 
and turns to increase below the onset temperature, $10$K, and then it saturates in the limit of $T\rightarrow 0$.
\cite{Tajima2006JPSJ}
The origin of such weak but obvious $T$-dependence has not been elucidated yet.
In the presence of magnetic field, $H$, perpendicular to the conducting plane, 
the two-step increase of resistivity is observed.\cite{Tajima2006JPSJ}
For example at $H=10$T, with decreasing $T$, the resistivity decreases weakly from room temperature and turns to increase below $T_0 \cong 20$K with the plateau in the lower temperature region.
Another relative sharp increase is observed around $T_l \cong 5$K leading to apparent saturation 
as $T\rightarrow 0$.
Both $T_0$ and $T_l$ increase with increasing magnetic field.

In the absence of tiling it has been established that the energy spectrum of the massless Dirac fermions 
becomes discrete by the Landau quantization owing to the orbital motion in the magnetic field, 
$E_N={\it sgn} (N) \sqrt{2 \hbar v^2 eH |N|/c}$, where $v$ is the velocity, and $N$ is an integer. 
Each Landau level has large degeneracy proportional to $H$, and is split into two states 
with up and down spins by the Zeeman energy, as shown in Fig. 1. 
The effects of tilting have been investigated recently and it is found that the Landau-level structure is 
qualitatively the same and that the effects of tilting are incorporated as modifications of effective velocity. 
By use of values of relevant parameters appropriate for $\alpha$-(BEDT-TTF)$_2$I$_3$,
\cite{Goerbig2008,Morinari2008} 
$E_1 \cong 5$meV and  $2E_Z \equiv g\mu_{\rm B}H \cong 1$meV with the g-factor $g=2$ for H=10T.
Hence we see that the energy scale seen in resistivity measurement, $T_0$ ($\cong 2$meV) and $T_l$, 
are smaller than $E_1$.
Thus, the observed two-step increase of resistivity in $\alpha$-(BEDT-TTF)$_2$I$_3$ may be attributed 
to the $N=0$ Landau levels, whose causes will be studied theoretically in this paper.

The long range Coulomb interaction plays an important role for massless Dirac fermions. 
The effective Coulomb interaction under magnetic field, $I$, is estimated as 
$I \cong e^2/\epsilon l_{\rm H} \cong 50 \sqrt{H{\rm [T]}}/\epsilon$ meV, 
where $l_{\rm H}$ is the magnetic length, $l_{\rm H} =\sqrt{\hbar c/eH}$.
Although the polarizability $\epsilon$ of $\alpha$-(BEDT-TTF)$_2$I$_3$ under high pressure 
has not been identified so far and then there is some ambiguity, 
it is demonstrated later that $N=0$ Landau states have instability toward the pseudo-spin ferromagnetism 
since the effective Coulomb interaction $I$ can exceed $2E_z$.
In the presence of tilting, it is shown that the electron correlation can give rise to 
the quantum Hall ferromagnet of the pseudo-spin  
(the degree of freedom on the valleys) with the help of the large degeneracy of the Landau levels.
The easy plane anisotropy of the pseudo-spin ferromagnet results from the back scattering processes  
which is the inter-valley scattering terms exchanging large momentum.
Moreover, it is shown that the effects of fluctuations of phase variables of the order parameters 
can be described by XY Heisenberg model leading to Kosterlitz-Thouless (KT) transition 
\cite{Kosterlitz-Thouless1972} at lower temperature.

\begin{figure}[htb]
\begin{center}
 \vspace{2mm}
 \leavevmode
\epsfysize=7cm\epsfbox{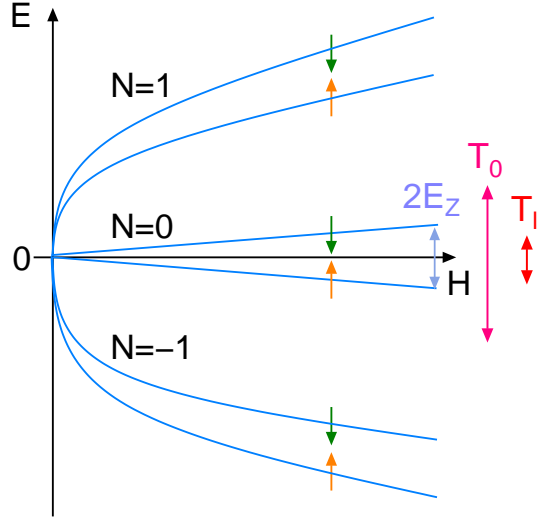}
 \vspace{-3mm}
\caption[]{Schematic figure of the Landau levels as a function of magnetic field.
The energy scale of the onset temperature of the first increase of resistivity, $T_0$, is located 
between the Zeeman gap $2E_Z$ and the $E_1 -E_0$ gap, where the Boltzmann factor $k_{\rm B}$ is taken as unit.
{\blue The energy scale of the second increase of resistivity at low temperatures, $T_l$, is smaller than the Zeeman gap.}
}
\label{fig1}
\end{center}
\end{figure}

\section{Formulation}

\subsection{Hamiltonian {\red for} massless Dirac fermions with tilting}
\label{Hamiltonian}

In the absence of magnetic field, the Hamiltonian of the massless Dirac fermions is given by,
{\magenta 
\begin{eqnarray}
{\cal H}&=&{\cal H}_0 +{\cal H}^\prime , \nonumber \\ 
{\cal H}_{\rm 0} &=&
\sum_{{\bf k} \gamma \gamma^\prime \sigma \tau} 
[ {\cal H}_0^{\sigma \tau} ]_{\gamma \gamma^\prime}
c^\dagger_{{\bf k} \gamma \sigma \tau} c_{{\bf k} \gamma^\prime \sigma \tau}  \nonumber \\ 
{\cal H}^\prime &=& \frac{1}{2} \int \int {\rm d}{\bf r} {\rm d}{\bf r}^\prime 
V_0 ({\bf r}-{\bf r}^\prime ) n({\bf r}) n({\bf r}^\prime )
\end{eqnarray}
} with the long-range Coulomb interaction $V_0 ({\bf r})= e^2 /\epsilon r$ and the density operator $n({\bf r})$.
{\magenta 
The degree of freedom on the spins are represented as $\sigma =\pm$ 
corresponding to $\uparrow$, $\downarrow$.
The degree of freedom of the pseudo-spins, $\tau =\pm$, corresponds to the valleys $R$, $L$. 
The valleys are located at the crossing points of the conduction and valence bands, 
$\pm {\bf k}_0$, where ${\bf k}_0$ is an incommensurate momentum in the first Brillouin zone.\cite{Katayama2006ZGS}
The creation and annihilation operators, 
$c^\dagger_{{\bf k} \gamma \sigma \tau}$ and $c_{{\bf k} \gamma \sigma \tau}$, respectively, 
are based on the Luttinger-Kohn representation\cite{LuttingerKohn} using the Bloch's functions 
at the crossing points as the basis of wave functions,\cite{Kobayashi2007} 
and then $\gamma =1,2$ denotes the basis of the Luttinger-Kohn representation.
The relation between the Luttinger-Kohn representation and the site representation based on the molecular orbitals 
are described in Appendix.}
The low-energy properties around the two crossing points, labeled by $\tau=\pm$, are described in 
terms of the two tilted Weyl Hamiltonians\cite{Kobayashi2007,Goerbig2008} 
\beqn\label{eq02}
&&H_0^{\sigma , \tau =+}=\hbar ( v \bk\cdot\sigmab + \bw_0\cdot \bk  \bone) \nonumber \\ 
&&H_0^{\sigma , \tau =-}=-\hbar ( v \bk\cdot\sigmab^* + \bw_0\cdot \bk  \bone) 
\eeqn
with respect to the time-reversal symmetry, where 
$v$ represents the velocity of the cone and $\bw_0$ represents the tilting velocity.
Here, we have neglected the anisotropy of the velocity of the cone.
In the case of the cone with anisotropic the velocity, 
one needs to replace $v \bk\rightarrow (v_x k_x, v_y k_y)$.

Once a magnetic field perpendicular to the conducting plane is taken into account, 
the momentum ${\bf k}$ is replaced by ${\bf k}+(e/c){\bf A}$ with the vector potential 
in the Landau gauge ${\bf A}=(0, Hx, 0)$, and then we obtain 
\beqn\label{eq02}
&&H_0^{\sigma , \tau =+}=\left[
\begin{array}{cc}
\hbar \{ w_{0x} \frac{1}{\rm i} \frac{\partial}{\partial x} + w_{0y} (\frac{1}{{\rm i}} \frac{\partial}{\partial y} 
+\frac{eHx}{c} ) \} -\sigma E_Z 
& \hbar v \{ \frac{1}{{\rm i}} \frac{\partial}{\partial x} 
-{\rm i} ( \frac{1}{{\rm i}} \frac{\partial}{\partial y} +\frac{eHx}{c} ) \}  \\
\hbar v \{ \frac{1}{{\rm i}} \frac{\partial}{\partial x} 
+{\rm i} ( \frac{1}{{\rm i}} \frac{\partial}{\partial y} +\frac{eHx}{c} ) \}
&  \hbar \{ w_{0x} \frac{1}{\rm i} \frac{\partial}{\partial x} 
+ w_{0y} (\frac{1}{{\rm i}} \frac{\partial}{\partial y} +\frac{eHx}{c} ) \} -\sigma E_Z \\         
\end{array}
\right] \nonumber \\
&&H_0^{\sigma , \tau =-}=\left[
\begin{array}{cc}
\hbar \{ w_{0x} \frac{1}{\rm i} \frac{\partial}{\partial x} 
+ w_{0y} (\frac{1}{{\rm i}} \frac{\partial}{\partial y} +\frac{eHx}{c} ) \} -\sigma E_Z 
& \hbar v \{ -\frac{1}{{\rm i}} \frac{\partial}{\partial x} 
-{\rm i} ( \frac{1}{{\rm i}} \frac{\partial}{\partial y} +\frac{eHx}{c} ) \} \\
\hbar v \{ -\frac{1}{{\rm i}} \frac{\partial}{\partial x} 
+{\rm i} ( \frac{1}{{\rm i}} \frac{\partial}{\partial y} +\frac{eHx}{c} ) \}
&  \hbar \{ w_{0x} \frac{1}{\rm i} \frac{\partial}{\partial x} 
+ w_{0y} (\frac{1}{{\rm i}} \frac{\partial}{\partial y} +\frac{eHx}{c} ) \} -\sigma E_Z \\         
\end{array}
\right]
\eeqn
in terms of the Zeeman energy $E_Z$.

\subsection{Zero-energy Landau level for the case with tilting}
\label{appC}

\YS{
The eigen equations for the zero-energy Landau level are given by 
\beq\label{WFT00}
H_0^{\sigma , \tau} \phi_X^{\tau}(\br) =0
\eeq
which gives the eigen functions of the tilted Weyl Hamiltonians in the presence of magnetic field.
The  wave functions are given by\cite{Goerbig2008} }
\beq\label{WFT}
\phi_X^{\tau}(\br)=\frac{1}{\sqrt{L}}e^{-{\rm i}X y/l_H^2}\varphi^{\tau}(x-X)  e^{-{\rm i}\tau {\bf k}_0\cdot {\bf r}}
\eeq
with 
\beq\label{WF+}
\varphi^\tau (x)=\frac{1}{\sqrt{\sqrt{\pi/\gamma}l_H}}
\chi_\tau
e^{-\gamma x^2/2l_H^2} ,
\eeq
where $X$ is the {\red guiding-center} coordinate and $L$ is the length of the system.
The spinor parts $\chi_\tau$ are given by 
\beqn
\label{spin+}
\chi_{\tau=+} &=& \frac{1}{\sqrt{\wtilde_0^2+(1+\gamma)^2}}\left(
\begin{array}{c} 
-\wtilde_0 e^{-i\varphi} \\ 1+\gamma          
\end{array}\right)\ , \\
\label{spin-}
\chi_{\tau=-} &=& \frac{1}{\sqrt{\wtilde_0^2+(1+\gamma)^2}}\left(
\begin{array}{c} 
 1+\gamma \\ -\wtilde_0 e^{+i\varphi}          
\end{array}\right)\ .
\eeqn
Here, we have defined
\beq
\wtilde_0 e^{i\varphi}\equiv \frac{w_{0x}+i w_{0y}}{v} ,
\eeq
in terms of the effective tilting parameter 
\beq\label{tiltparam}
\wtilde_0\equiv\sqrt{\left(\frac{w_{0x}}{v}\right)^2+
\left(\frac{w_{0y}}{v}\right)^2}
\eeq
with $\bw_0 =(w_{0x},w_{0y})$, and $\gamma=\sqrt{1-\wtilde_0^2}$.
We note that one recovers the usual result for the $n=0$ wave function in graphene 
when we use in the limit $\wtilde_0\rightarrow 0$
($\gamma\rightarrow 1$), i.e. in the case without tilting.

\subsection{Effective Hamiltonian on fictitious magnetic lattice}

We consider the present model by using the bases of the Wannier functions 
for the magnetic rectangular lattice which is introduced fictitiously.
The wave functions in the Landau gauge 
are not localized in the $y$-direction, {\red but in the $x$-direction around the position $X$.}
By applying the periodic boundary condition in the $y$-direction with the system length $L$, 
the $X$ is discretized as $X=-2 \pi l_{\rm H}^2 j /L$ with an integer $j$.
The Wannier functions, which satisfy orthonormality and are localized 
around ${\bf R} =(ma,nb)$ with integers $m$ and $n$ 
as shown in Fig. 2, are constructed by the linear combination of $\phi_{X} ({\bf r})$,\cite{Fukuyama1977} 
\beq\label{wannier}
\Phi_{\bR}^{\tau}(\br)=\frac{\sqrt{L}}{a\sqrt{b}}\int_{-a/2}^{a/2} dX\, e^{i X nb/l_H^2}\phi_{X+ma}^{\tau}(\br)\ ,
\eeq
where $a$ is arbitrary but $a \gg 2 \pi l_{\rm H}^2 /L$ and  $b=2 \pi l_{\rm H}^2 /a$. 
We note that $\vert \Phi_{\bR}^{\tau}(\br) \vert$ exhibits an exponential-like decrease in the $x$-direction, but decreases 
{\red algebraically} as $\vert y \vert^{-1}$ in the $y$-direction.

\begin{figure}[htb]
\begin{center}
 \vspace{2mm}
 \leavevmode
\epsfysize=7cm\epsfbox{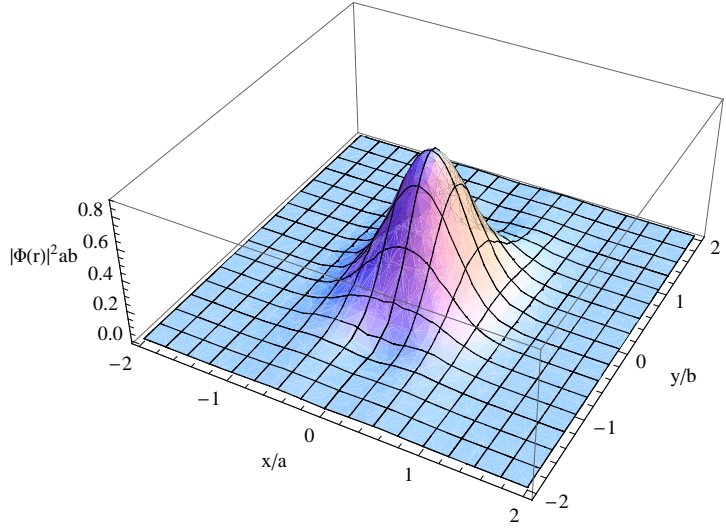}
 \vspace{-3mm}
\caption[]{
The ${\bf r}$-dependence of $\vert \Phi_{\bR}^{\tau}(\br) \vert^2$ with ${\bf R}={\bf 0}$, 
where $\Phi_{\bR}^{\tau}(\br)$ is the Wannier functions for the magnetic rectangular lattice.
}
\label{fig2}
\end{center}
\end{figure}

In the basis of these Wannier functions, the effective interaction Hamiltonian is given by
\begin{eqnarray}
&& {\cal H}_{N=0}^\prime = \frac{1}{2} \sum_{{\bf R}_{1,2,3,4}} 
\sum_{\sigma \sigma^\prime \tau_{1,2,3,4}} 
\int \int {\rm d}{\bf r} {\rm d}{\bf r}^\prime V_0 ({\bf r}-{\bf r}^\prime ) \nonumber \\
&& \times {\red \left[ \Phi_{{\bf R}_1}^{\tau_1} ({\bf r})^\dag \cdot \Phi_{{\bf R}_2}^{\tau_2} ({\bf r}) \right] 
\left[ \Phi_{{\bf R}_3}^{\tau_3} ({\bf r}^\prime)^\dag \cdot \Phi_{{\bf R}_4}^{\tau_4} ({\bf r}^\prime) \right]} \nonumber \\
&& \times 
c_{{\bf R}_1 \sigma \tau_1}^\dagger c_{{\bf R}_2 \sigma \tau_2} 
c_{{\bf R}_3 \sigma^\prime \tau_3}^\dagger c_{{\bf R}_4 \sigma^\prime \tau_4} 
\end{eqnarray}
under the assumption that the density operator is effectively determined by the field operator 
for $N=0$ Landau states and the contributions to the Hamiltonian from $N \ne 0$ states are negligible, 
{\it i. e.} we use the density operator
\begin{eqnarray}
n ({\bf r})= \sum_{{\bf R} {\bf R}^\prime \sigma \tau \tau^\prime} 
 {\red \left[ \Phi_{\bf R}^\tau ({\bf r})^\dag \cdot  \Phi_{{\bf R}^\prime}^{\tau^\prime} ({\bf r}) \right]}
 c_{{\bf R} \sigma \tau}^\dagger  c_{{\bf R}^\prime \sigma \tau^\prime}.
\end{eqnarray}

Thus the effective Hamiltonian for the $N=0$ Landau states in the magnetic rectangular lattice is given by 
\begin{eqnarray}
{\cal H}_{\rm eff} &=& 
\sum_{i \sigma \tau} (-\sigma E_Z) c^\dagger_{i \sigma \tau} c_{i \sigma \tau} \nonumber \\
&+& \sum_{i j k l \sigma \sigma^\prime \tau \tau^\prime} V_{i j k l} 
c^\dagger_{i \sigma \tau} c_{j \sigma \tau} 
c^\dagger_{k \sigma^\prime \tau^\prime} c_{l \sigma^\prime \tau^\prime}  \nonumber \\
&+& \sum_{i j k l \sigma \sigma^\prime \tau} W_{i j k l} 
c^\dagger_{i \sigma \bar{\tau}} c_{j \sigma \tau} 
c^\dagger_{k \sigma^\prime \tau} c_{l \sigma^\prime \bar{\tau}} ,
\end{eqnarray}
where $i$, $j$, $k$, and $l$ denote the unit cells of the magnetic rectangular lattice 
at ${\bf R}_i$, ${\bf R}_j$, ${\bf R}_k$, and ${\bf R}_l$, respectively, and $\bar{\tau} =-\tau$.

The forward-scattering term, $V_{i j k l}$, is given by
\begin{equation}
V_{i j k l} = \frac{1}{2} \int \int {\rm d}{\bf r} {\rm d}{\bf r}^\prime \,
V_0 ({\bf r}-{\bf r}^\prime ) \,
{\red \left[ \Phi_i^\tau ({\bf r})^\dag \cdot \Phi_j^\tau ({\bf r}) \right] 
\left[ \Phi_k^{\tau^\prime} ({\bf r}^\prime)^\dag \cdot \Phi_l^{\tau^\prime} ({\bf r}^\prime) \right]}
\end{equation}
from the long wave length part of ${\cal H}_{N=0}^\prime$.
This term does not depend on the spin and pseudo-spin, and then 
it does not break the SU(4) symmetry, {\red neither in the spin subspace nor in that of the pseudo-spin}.
We find that the forward-scattering term is not affected by the tilting, 
because $( \chi_{\tau}^{\dag} \cdot \chi_{\tau} )( \chi_{\tau^\prime}^{\dagger} \cdot \chi_{\tau^\prime} )=1$.

On the other hand, the backscattering term, $W_{i j k l}$, 
which is the inter-valley scattering term exchanging large momentum $2k_0$ and 
breaks the SU(2) symmetry in the subspace of the pseudo-spin, 
is given by
\begin{equation}
W_{i j k l} = \frac{1}{2} \int \int {\rm d}{\bf r} {\rm d}{\bf r}^\prime \,
V_0 ({\bf r}-{\bf r}^\prime ) \,
\red{\left[ \Phi_i^{\bar{\tau}} ({\bf r})^\dag \cdot \Phi_j^\tau ({\bf r}) \right]  
\left[ \Phi_k^\tau ({\bf r}^\prime)^\dag \cdot \Phi_l^{\bar{\tau}} ({\bf r}^\prime) \right]}
\end{equation}
from the short wave length part of ${\cal H}_{N=0}^\prime$.
In the absence of tilting, {\red as e.g. in graphene,} the backscattering term vanishes because 
$( \chi_{\bar{\tau}}^{\dag} \cdot \chi_{\tau} )=0$.\cite{Goerbig2006}
We find that the tilting is essential to have a non-zero backscattering term.
The tilting dependence of the backscattering term is given by the spinor part, 
\beq\label{BW}
( \chi_{\bar{\tau}}^{\dag} \cdot \chi_{\tau} )( \chi_{\tau}^{\dagger} \cdot \chi_{\bar{\tau}} )
=\frac{4\wtilde_0^2(1+\gamma)^2 }
{[\wtilde_0^2+(1+\gamma)^2]^2}\simeq 4\wtilde_0^2 + O (\wtilde_0^4) ,
\eeq
where the last step has been obtained from the limit $\wtilde_0\ll 1$.
{\magenta 
The ratio between the forward and the backscattering terms, $W_{ijkl}/V_{ijkl}$, is given by 
\beq\label{BW}
W_{ijkl}/V_{ijkl} 
\simeq \frac{\wtilde_0^2 a_{\rm L}}{l_{\rm H}} ,
\eeq
}where $a_{\rm L}$ is the lattice constant in the conducting plane.
The backscattering term is proportional to $a_{\rm L}$, 
since the large momentum $\vert 2 k_0 \vert \cong \pi /a_{\rm L}$ is exchanged.\cite{Goerbig2006}
We note that the lattice constant of $\alpha$-(BEDT-TTF)$_2$I$_3$, $a_{\rm L} \cong 10 {\rm \AA}$, 
is much larger than that of graphene.
Thus it is expected that the backscattering term {\red plays an important role for electron-correlation effects 
in $\alpha$-(BEDT-TTF)$_2$I$_3$}.
The typical value of the ratio $W_{ijkl}/V_{ijkl}$ is approximately $0.07$ for $\alpha$-(BEDT-TTF)$_2$I$_3$ 
at $H=10$T using the tilting parameter $\wtilde_0 \cong 0.8$.
The Umklapp scattering term ($\tau \bar{\tau} \tau \bar{\tau}$-term) can be neglected, 
because it is exponentially smaller than the other terms as a function of $a_{\rm L}/l_{\rm H}$, 
which is estimated as $0.1$ at $H=10T$ in $\alpha$-(BEDT-TTF)$_2$I$_3$.

\section{Pseudo-spin ferromagnet and Kosterlitz-Thouless transition}~

Possible spin and pseudo-spin ferromagnetic states in the zero-energy Landau level in graphene have
been extensively studied in recent years.\cite{Goerbig2006,NomuraMacDonald2006,Alicea2006,Yang2006,Gusynin2006,Herbut2007,doretto,Ezawa2007}
Generically, the ferromagnetic ordering may be understood within an interaction model with no explicit
spin or pseudo-spin symmetry breaking; in order to minimize their exchange energy, the global $N$-particle 
wave function should be fully antisymmetric in its orbital part, the (pseudo-)spin part needs to be fully
symmetric in order to fulfil fermionic statistics. Whereas in a normal metal this ordering is only partial, due
to the increase in the kinetic energy, a single Landau level may be viewed as an infinitely flat energy band,
and the ferromagnetic ordering may therefore be complete. In the absence of an explicit symmetry breaking, such as
the Zeeman effect that naturally tends to polarize the physical spin or the above-mentioned backscattering term
that affects the pseudo-spin, no particular spin or pseudo-spin channel is selected, and one may even find 
an entangled spin-pseudo-spin ferromagnetic state.\cite{Doucot2008} The symmetry-breaking terms may, thus, be
viewed as ones that choose a particular channel (spin or pseudo-spin) and direction of a pre-existing ferromagnetic
state by explicitly breaking the original SU(4) symmetry.

\subsection{Mean-field solution}

The mean-field Hamiltonian for the pseudo-spin ferromagnetic state is given by
\begin{eqnarray}
&& {\cal H}_{\rm MF} = \sum_{j \sigma \tau} \left[-\sigma E_Z -2\sum_i 
(V_{ijji} n_{i\sigma \tau} +W_{ijji} n_{i\sigma \bar{\tau}}) \right]
c_{j\sigma \tau}^\dagger c_{j\sigma \tau} \nonumber \\
&& -2\sum_{ij\sigma \tau} (V_{ijji} +W_{iijj}) (\langle c_{i\sigma \tau}^\dagger c_{i\sigma \bar{\tau}} \rangle  
c_{j\sigma \bar{\tau}}^\dagger c_{j\sigma \tau} +{\rm h.c})
\end{eqnarray}
and the order parameter of the pseudo-spin ferromagnetic state, $\Delta$, 
which is independent of $i$ and $\sigma$, is defined by 
\begin{equation}
\Delta = 2I \langle c_{i\sigma -}^\dagger c_{i\sigma +} \rangle 
\end{equation}
with the effective interaction $I=\sum_i (V_{i00i} +W_{ii00})$.
Using the spin polarization, $m = \sum_{\sigma \tau} \sigma n_{i\sigma \tau}$, which is also independent of $i$, 
and the renormalized Zeeman energy, $\tilde{E}_Z = E_Z +mI$, the mean-field Hamiltonian is given by 
\begin{eqnarray}
&& {\cal H}_{\rm MF} = \sum_{j} {\bf c}_j^\dagger \hat{{\cal H}} {\bf c}_j \nonumber \\ 
&& \hat{{\cal H}} = \left[
\begin{array}{cccc}
-\tilde{E}_Z & -\Delta^\ast & 0 & 0 \nonumber \\ 
-\Delta & -\tilde{E}_Z & 0 & 0 \nonumber \\ 
0 & 0 & \tilde{E}_Z & -\Delta^\ast \nonumber \\ 
0 & 0 & -\Delta & \tilde{E}_Z \nonumber \\ 
\end{array}
\right]
\end{eqnarray}
with ${\bf c}_j =(c_{j \uparrow +}, c_{j \uparrow -}, c_{j \downarrow +}, c_{j \downarrow -} )$.
The mean-field solution is calculated from  
\begin{equation}
\vert \Delta \vert =\frac{2I}{\pi} \int {\rm d}x f(x) {\rm Im} 
\frac{\vert \Delta \vert}{(x+{\rm i}\delta -\sigma \tilde{E}_Z)^2 -\vert \Delta \vert^2} 
\end{equation}
and 
\begin{equation}
m =\frac{1}{2} \sum_{\sigma \tau} \sigma f(-\sigma \tilde{E}_Z +\tau \vert \Delta \vert )
\end{equation}
with the Fermi distribution function, $f(x)$. 

The ground state in the case with $I < E_Z$ is a spin polarized state 
without pseudo-spin polarization ($m=1$ and $\Delta =0$ at $T=0$), 
where electrons reside in the spin-up branches of the $N=0$ Landau levels 
(see the left hand side of fig. 3(a)).
The ground state in the case with $I > E_Z$ is, on the other hand, 
a pseudo-spin ferromagnetic state ($m=0$ and $\vert \Delta \vert =I$ at $T=0$), 
where the easy-plane pseudo-spin polarization lifts 
the pseudo-spin degeneracy and then the spin polarization is suppressed 
(see the right hand side of fig. 3(a)).

The mean-field phase diagram in the $I$-$T$ plane scaled by $E_Z$ is shown in fig. 3(b).
The transition temperature for the easy-plane pseudo-spin ferromagnetic state, $T_{\rm c}$, 
is finite in the case with $I > E_Z$, and increases with increasing $I$.
Below $T_{\rm c}$, the spin polarization vanishes at $T \rightarrow 0$, 
although it is still finite at finite temperatures.

One notices that this competition between a spin polarized state and an easy-plane pseudo-spin ferromagnetism 
is original to the filling factor $\nu=0$, 
where necessarily two (of four) subbranches of the zero-energy Landau levels are occupied. In contrast
to this particular filling factor, this competition is absent at $\nu=\pm 1$, 
where only one subbranch is occupied and where,
therefore, a spin polarization does not exclude a simultaneous pseudo-spin ordering 
in a coherent superposition of both pseudo-spin states.

\begin{figure}[htb]
\begin{center}
 \vspace{2mm}
 \leavevmode
\epsfysize=7cm\epsfbox{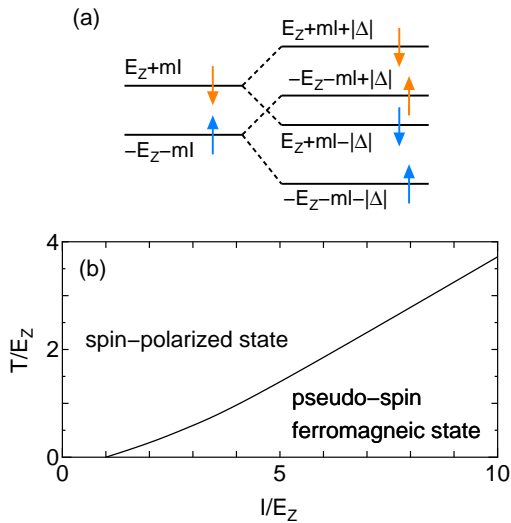}
 \vspace{-3mm}
\caption[]{
(a) Schematic figure of the energy levels in the spin-polarized state (left hand side) 
and the pseudo-spin ferromagnetic state (right hand side).
(b) Phase diagram in the plane of the interaction $I$ and temperature $T$ scaled by the Zeeman energy $E_Z$.
}
\label{fig3}
\end{center}
\end{figure}

\subsection{Phase fluctuations and Kosterlitz-Thouless transition}

In the presence of {\red an order parameter with a finite amplitude} below $T_c$,
phase fluctuation exists with the characteristic length of spatial variation 
much longer than the fictitious lattice spacing.
The effect of these phase fluctuation, {\red which has so far been ignored} in the mean-field approximation, is treated 
on the basis of the Wannier functions and the resulting model is similar to the {\red XY model}
leading to the KT transition.
Using the pseudo-spin operator, $\tilde{S}_{i-}^\sigma \equiv c_{i\sigma L}^\dagger c_{i\sigma R}$, 
and the real spin operator, $S_{jz} =\frac{1}{2} \sum_{\sigma \tau} \sigma c_{j\sigma \tau}^\dagger c_{j\sigma \tau}$, 
the mean-field Hamiltonian is given by 
\begin{eqnarray}
{\cal H}_{\rm MF} &=& -2(E_Z + mI) \sum_j S_{jz} \nonumber \\
&-& 2\sum_{ij\sigma} I_{ij} \left(\langle \tilde{S}_{i-}^\sigma \rangle  
\tilde{S}_{j+}^\sigma + \langle \tilde{S}_{i+}^\sigma \rangle  
\tilde{S}_{j-}^\sigma \right) .
\end{eqnarray}
The real spin polarization, $m$, remains finite at finite temperatures, 
although it vanishes at $T \rightarrow 0$ in the pseudo-spin ferromagnetic state.
However, it can be shown that the spin polarization is independent of $i$, 
since the pseudo-spin can fluctuate only in the easy plane and then 
the occupation numbers of electrons are independent of $i$.
The interactions between the pseudo-spins on the magnetic rectangular lattice, 
$I_{ij} \equiv V_{ijji}+W_{iijj}$.
The interaction $I_{ij}$ rapidly decreases with increasing $\vert {\bf R}_j -{\bf R}_i \vert$.
It is numerically found {\blue as seen in Fig. 4} that the nearest-neighbor $I_{ij}$ 
is approximately isotropic and the ratio 
$I_{i,i+1} /I_{i,i}=0.10$ with the arbitrary choice of $b=\sqrt{2}a$ 
(but leading to the almost isotropic localization of the Wannier function on the fictitious lattice), 
\YS{ where
\beq\label{AD00}
\frac{I_{i,j}}{I_{i,i}} = \frac{V_{ijji}+W_{iijj}}{V_{iiii}+W_{iiii}} \cong \frac{V_{ijji}}{V_{iiii}}
\eeq }
The phase of the order parameter corresponds to the angles of the pseudo-spin.
The $x$- and $y$-components of the pseudo-spins are given by 
$\langle \tilde{S}_{ix}^\sigma \rangle ={\rm Re} \langle \tilde{S}_{i-}^\sigma \rangle$ and  
$\langle \tilde{S}_{iy}^\sigma \rangle =-{\rm Im} \langle \tilde{S}_{i-}^\sigma \rangle$, respectively,  
with $\langle \tilde{S}_{i+}^\sigma \rangle =\langle \tilde{S}_{i-}^\sigma \rangle^\ast$. 
Thus $\langle \tilde{S}_{i-}^\sigma \rangle$ can be represented using the angle of the pseudo-spin from the $x$-direction 
in the $x$-$y$ plane, $\phi_i^\sigma$, as 
$\langle \tilde{S}_{i-}^\sigma \rangle = \vert \langle \tilde{S}_{i-}^\sigma \rangle \vert 
\exp (-{\rm i}\phi_i^\sigma ) $.
When the characteristic length of spatial variation of the phases is much longer than the lattice spacing, 
we can expand the free energy by the fluctuations of the phases 
$f_{ij}^\sigma =1-\cos (\phi_j^\sigma -\phi_i^\sigma ) \ll 1$ 
under the assumption that the amplitude, $\vert \langle \tilde{S}_{i-}^\sigma \rangle \vert$, 
does not change within the characteristic length of the phase fluctuation, and then we obtain 
\begin{eqnarray}
F_f -F_n & \cong & (F_f -F_n )\vert_{f_{ij}^\sigma=0} +\sum_{\langle i \ne j \rangle \sigma} 
\frac{\partial (F_f -F_n )}{\partial f_{ij}^\sigma} f_{ij}^\sigma \nonumber \\
&=& F_0 -\sum_{\langle i \ne j \rangle \sigma} J_{ij}^\sigma \cos (\phi_j^\sigma -\phi_i^\sigma )
\end{eqnarray}
with $J_{ij}^\sigma =4 \vert \langle \tilde{S}_{-}^\sigma \rangle \vert^2 I_{ij}$, 
where $F_f$ and $F_n$ denote the free energies of the pseudo-spin ferromagnetic and normal states, respectively, 
and $F_0$ is independent of the phases.
The effects of the phase fluctuations in $J_{ij}^\sigma$ are neglected since those effects are the higher-order terms, 
and $J_{ij}^\sigma \rightarrow I_{ij}$ at temperatures much lower than $T_{\rm c}$.
Then the physics of the phase fluctuation is equivalent to that of the two-dimensional $XY$ Heisenberg model 
with the nearest-neighbor exchange interaction, $J \cong I_{i,i+1} \cong 0.087 I$. 
{\blue In the two-dimensional XY model, the KT transition occurs 
due to the onset of bound pairs of the vortices.\cite{Kosterlitz-Thouless1972}}
It is known that $T_{\rm KT} \cong 1.54 J$ by the renormalization group analysis.\cite{Kosterlitz1974}
Here the effects of spin polarization are negligible, because 
$J$ is much larger than the Zeeman energy and the phase fluctuation on each electron with up or down spin 
is described by the same interaction, $J$.

Lastly we discuss the role of the long-range part of $I_{ij}$ farther than the nearest neighbor one.
Figure 4 shows the distance dependences of $I_{0,j}/I_{0,0}$ with ${\bf R}_j =(na,0)$ (the closed circles) and 
${\bf R}_i =(0,nb)$ (the open circles) defined on an integer $n$, where we take $b=\sqrt{2}a$.
The interaction $I_{0j}$ decays very rapidly along the $x$-axis but slowly along $y$-axis.
Then the role of the long range part of $I_{ij}$ along $y$-axis should be considered for improving 
the effective Hamiltonian. 
The KT transition temperature $T_{\rm KT}$, however, is determined by the competition 
between the excitation energy of a vortex and the entropy effect coming from the degree of freedom 
for the position of the vortex core.
Since the length scale of the vortex is much longer than $a$ and 
the interaction is ferromagnetic, 
the long-range part of $I_{ij}$ does not disturb the KT transition essentially. 

\begin{figure}[htb]
\begin{center}
 \vspace{2mm}
 \leavevmode
\epsfysize=7cm\epsfbox{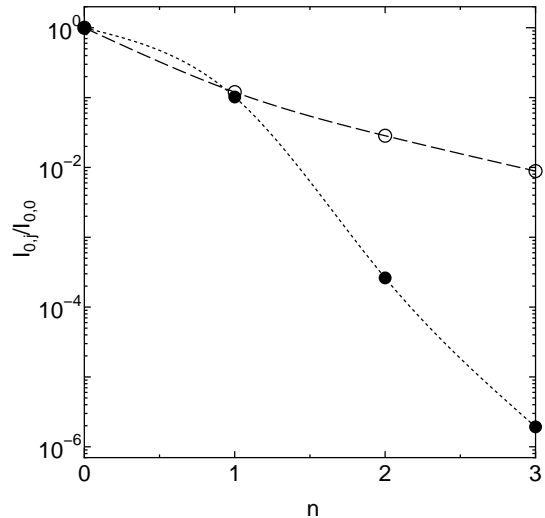}
 \vspace{-3mm}
\caption[]{
The $n$-dependences of $I_{0,j}/I_{0,0}$ with ${\bf R}_j =(na,0)$ (the closed circles) and 
${\bf R}_i =(0,nb)$ (the open circles), where we take $b=\sqrt{2}a$.
The dashed and dotted lines are {\red a guide to the eye}.
}
\label{fig4}
\end{center}
\end{figure}

\section{Relation between Experimental Findings and Theoretical Results}

In the presence of magnetic field perpendicular to the conducting plane, 
the two-step increase of resistivity, for example at $T_0 \cong 20$K and $T_l \cong 5$K at $H=10$T, 
is observed in $\alpha$-(BEDT-TTF)$_2$I$_3$, 
where both $T_0$ and $T_l$ increase with increasing magnetic field.\cite{Tajima2006JPSJ}
We may be able to associate two stepwise changes of resistivity 
as due to the easy-plane pseudo-spin ferromagnetic transition at $T_{\rm c}$ and the KT transition at $T_{\rm KT}$, 
since our estimate indicates that $T_{\rm c} \cong 4 T_{\rm KT}$, 
where $T_{\rm c} \cong 0.5 I$ as seen in Fig. 3(b) in the region of $I/E_Z >>1$ of interest and 
$T_{\rm KT} \cong 1.54J \cong 0.13 I$.
We emphasize that the tilting of the Dirac cone is essential to the appearance of the easy-plane 
pseudo-spin ferromagnet, and thus, 
to the appearance of the KT transition, due to the long range Coulomb interaction.

\section{Conclusion and Discussion}

In the present paper, motivated by the experimental observation of the particular temperature dependences 
of resistivity in $\alpha$-(BEDT-TTF)$_2$I$_3$ under magnetic field, the
possibility of the pseudo-spin quantum Hall ferromagnet at $\nu =0$ has been investigated 
in the massless Dirac fermion system. 
The pseudo-spin ferromagnetic transition occurs when the electron correlation exceeds the Zeeman energy.
The tilting of the Dirac cone induces the backscattering terms resulting in the easy-plane pseudo-spin ferromagnet.
There will be intrinsic fluctuations and $T_{\rm c}$ should be considered 
only as a crossover temperature for the growing amplitude of order parameters with remaining large phase fluctuations 
in the two-dimensional system.
To treat such phase fluctuations, {\red a} spatially localized basis set similar to ``Wannier function'' 
are introduced, which indicates that the model is similar to {\red the XY model} which is known to lead to 
the KT transition at lower temperature, $T_{\rm KT}$.
In comparison with experiments, the two-step increase of resistivity with decreasing temperature are observed 
 at around $20$K and $5$K at $H= 10$T. 
Present theory has revealed $T_{\rm KT} \cong 5$K on the choice of the parameters giving $T_{\rm c} \cong 20$K, 
and then there are reasonable correspondences to identify two stepwise changes of resistivity 
as due to the amplitude growing and the phase coherence of the order parameters.

Obviously there are remaining problems to be clarified. 
The experimental data indicates the saturation of resistivity in the low temperature. 
The saturation indicates the existence of dilute carriers which may originate from 
weak three-dimensionality or disorder. 

In graphene,\cite{graphene} the the quantum Hall ferromagnet at $\nu =0$ has been investigated,
\cite{Goerbig2006,Alicea2006,NomuraMacDonald2006,Gusynin2006,Ezawa2007} 
and very recently it is suggested that the electron-phonon interaction breaking the pseudo-spin SU(2) symmetry, 
which may be characteristic of graphene, induces the easy-plane pseudo-spin ferromagnet 
resulting in the KT transition \cite{Nomura2009} in order to explain the possible KT transition 
observed in graphene.\cite{Ong2008} 
We emphasize that the backscattering term, which is the key factor for the easy-plane pseudo-spin ferromagnet 
in our paper, is characteristic of $\alpha$-(BEDT-TTF)$_2$I$_3$, but can be realized in graphene 
by distorting the honeycomb lattice.
In addition, we note that a lattice model describing the fluctuations of the pseudo-spins 
should be based on Wannier functions which satisfy orthonormality under magnetic field, 
since the bases on the original crystal lattice are no longer the eigenstates of the $N=0$ Landau levels.

The effects of the short range parts of the Coulomb interaction in graphene also have been investigated.
If once the pseudo-spin ferromagnetism occurs, the Hubbard-$U$-type on-site interaction favors 
the easy-plane ferromagnetism (uniform charge density), 
while the nearest-neighbor interaction $V$ favors the easy-axis ferromagnetism 
(the sublatice CDW), in the case of graphene.\cite{Alicea2006}
In $\alpha$-(BEDT-TTF)$_2$I$_3$, however, the easy-axis ferromagnetism does not 
correspond to CDW directly, because the bases of the Weyl Hamiltonian is not the sublattice 
but the Bloch states at ${\bf k}=\pm {\bf k}_0$. 
The Bloch states at ${\bf k}=\pm {\bf k}_0$ consist of the linear combination 
of the contributions from four BEDT-TTF molecules. 
Thus, although the easy-axis ferromagnetism may modify 
the intrinsic charge disproportionation in $\alpha$-(BEDT-TTF)$_2$I$_3$, 
the effect of $V$ on such state may be weaker than than of graphene.
It is an interesting difference between graphene and $\alpha$-(BEDT-TTF)$_2$I$_3$, 
and it will be investigated intensively in future.

Lastly, we discuss the renormalization of fluctuation in the pseudo-spin ferromagnet, 
which are very complicated and not captured on the mean-field level. 
In the absence of the Zeeman effect, 
the ferromagnetic moment may fluctuate in the SU(4) space at temperatures 
between $T_{\rm c}$ and a symmetry breaking temperature, $T_{\rm sb}$, which is essentially given by 
the symmetry-breaking interaction energy, $W_{ijkl}$.
The ferromagnetic moment may be forced in the easy-plane below $T_{\rm sb}$.
In the presence of the Zeeman effect, the situation is more complicated 
owing to the competition between $W_{ijkl}$ and the Zeeman energy, $E_Z$.
The results in the present paper, thus, identify a new research target, {\it i. e.} 
a two-dimensional SU(4) model with the symmetry-breaking terms.

\section*{ Acknowledgments }

The authors are thankful to N. Tajima for fruitful discussions.
This work has been financially supported by Grant-in-Aid for 
Special Coordination Funds for Promoting Science and Technology (SCF), 
Scientific Research on Innovative Areas 20110002, and 
Scientific Research 19740205 
from the Ministry of Education, Culture, Sports, Science and Technology in Japan.



\magenta 

\appendix

\section{Relation between Luttinger-Kohn Representation and Site Representation}

A basic model describing the two-dimensional electronic system in $\alpha$-(BEDT-TTF)$_2$I$_3$ 
is given by\cite{Mori1984,SeoRev2004,Kobayashi2004}
\begin{equation}
{\cal H}_{\rm site} = \sum_{( i \alpha : j \beta ), \sigma}
 (t_{i \alpha; j \beta}\ 
   a^{\dag}_{i\alpha\sigma}a_{j\beta\sigma}+ {\rm h. c.} ) ,
\end{equation}
where $i$ and $j$ denote the unit cells, $\alpha$ ,$\beta$ 
(=A, A', B and C) denote the molecular orbital sites in a unit cell, 
$a^\dagger_{i\alpha \sigma }$ denote the creation operators on the site representation, 
and $t_{i\alpha ; j \beta }$ is the transfer energy between $(i,\alpha )$ site and $(j,\beta )$ site.
Using the Fourier transformation, we obtain 
\begin{eqnarray}
&& {\cal H}_{\rm site}
 = \sum_{{\bf k}\alpha\beta\sigma}{}
\epsilon_{\alpha\beta\sigma}({\bf k}) 
 a^{\dag}_{{\bf k}\alpha\sigma}a_{{\bf k}\beta\sigma}\nonumber\\
&& \epsilon_{\alpha\beta}({\bf k}) = \sum_{{\bf \delta}}
t_{\alpha\beta} e^{{\rm i} {\bf k}\cdot {\bf \delta}},
\end{eqnarray}
where ${\rm \delta}$ denotes the vector representing the nearest neighbor of the unit cell.
Such Hamiltonian (here we call it the site Hamiltonian) is diagonalized by the eigenvalue equation  
\begin{eqnarray}
\sum_{\beta=1}^{4}
\epsilon_{\alpha\beta\sigma}({\bf k})\ d_{\beta \gamma \sigma}({\bf k})
&=&\xi_{\gamma \sigma}({\bf k})\ d_{\alpha \gamma \sigma}({\bf k}) ,
\label{eigenvalue}
\end{eqnarray}
where $\xi_{\gamma \sigma}$ are the eigenvalue with the descending order, 
$\xi_{1 \sigma}({\bf k}) > \xi_{2 \sigma}({\bf k}) 
  > \xi_{3 \sigma}({\bf k}) > \xi_{4 \sigma}({\bf k})$, 
and $d_{\alpha r \sigma}({\bf k})$ ($\gamma =1,2,3,4$) are the corresponding eigenvectors.
Here $\xi_{1 \sigma}({\bf k})$ and $\xi_{2\sigma}({\bf k})$ are the conduction and valence bands, respectively, 
because there are six electrons in the four molecules, i.e., the $3/4$-filled electronic system.

Expanding the site Hamiltonian in the linear order of momenta from $\pm {\bf k}_0^\prime$ 
and using the Luttinger-Kohn representation based on the Bloch's functions at $\pm {\bf k}_0^\prime$ 
(corresponding to $\tau =\pm$), we obtain the tilted Weyl Hamiltonians, 
where $\pm {\bf k}_0^\prime$ are infinitesimally close to the crossing points $\pm {\bf k}_0$, 
respectively.\cite{Kobayashi2007,Kobayashi2009Rev} 
Then the creation operators on the Luttinger-Kohn representation 
$c^{\dagger}_{{\bf k} \gamma \sigma \tau}$ are given by
\begin{equation}
c^{\dagger}_{{\bf k} \gamma \sigma \tau} =\sum_\alpha d_{\alpha \gamma \sigma}^{\ast} (\tau {\bf k}_0^\prime ) 
a^{\dagger}_{{\bf k} \alpha \sigma}.
\end{equation}



\end{document}